\documentclass[11pt]{article}
\usepackage[utf8]{inputenc}

\usepackage{graphicx}
\usepackage{hyperref}
\usepackage[round,semicolon,authoryear]{natbib}
\usepackage{amsmath}
\usepackage{xfrac}
\usepackage{algorithm}
\usepackage{algpseudocode}
\usepackage{authblk}
\usepackage{abstract}
\usepackage{booktabs}
\usepackage{multirow}
\usepackage{array}
\usepackage{rotating}
\usepackage{subcaption} 
\usepackage{xcolor}
\usepackage{mathtools}
\usepackage{bm}
\usepackage{arydshln}
\usepackage{bbm}
\usepackage{float}
\usepackage{pgfplotstable} 

\usepackage{amssymb}
\usepackage{upgreek}
\usepackage{tikz}
\graphicspath{{./Graphs/}}

\usetikzlibrary{arrows,automata}
\usetikzlibrary{positioning}
\usetikzlibrary{shapes}

\tikzset{
	>=stealth',
	arrow/.style={thick, ->, >=stealth},
	punkt/.style={
		ellipse,
		rounded corners,
		draw=black,
		text width=2.5em,
		minimum height=2em,
		text centered},
	pil/.style={
		->,
		shorten <=2pt,
		shorten >=2pt,},
	pil1/.style={ellipse,draw = white}
}

\providecommand{\keywords}[1]{%
  {\small\noindent\textbf{\textit{Keywords---}}#1\par}%
}
\providecommand{\msccodes}[1]{%
  {\small\noindent\textbf{\textit{MSC2020 codes---}}#1\par}%
}
\providecommand{\jelcodes}[1]{%
  {\small\noindent\textbf{\textit{JEL codes---}}#1\par}%
}
\usepackage[letterpaper, margin=1in]{geometry}
\usepackage{setspace}
\author[a,*]{Samuel Modée}
\author[b]{Yushu Li}
\author[c]{Sjur Westgaard}
\author[d]{Stein Andreas Bethuelsen}

\affil[a,b,d]{Department of Mathematics, University of Bergen, Norway}

\affil[c]{Department of Industrial Economics and Technology Management,
	NTNU, Trondheim, Norway}
\affil[*]{Corresponding author. E-mail: \texttt{samuel.modee@uib.no}}

\pgfplotsset{compat=1.18}

\begin{document}

\title{Multi-regime Markov-switching models with time-varying transition probabilities: An application to U.S. Treasury yields}
\date{}
\maketitle

\begin{abstract}
\noindent This paper studies Markov-switching (MS) models with time-varying transition probabilities (TVTP) under alternative specifications of the transition probability matrix. We extend the two-regime common-variance setting of the Generalized Autoregressive Score (GAS) model of \citet{bazzi2017time} to the general $K$-regime case with regime-specific means and variances, and develop an open-source R package, \texttt{multiregimeTVTP}, for simulation and estimation of such models. A Monte Carlo study shows that the regime means, variances, and transition probabilities are reliably recovered, whereas the TVTP driving coefficients are harder to identify. The GAS score coefficient appears to be statistically non-identifiable, owing to a ridge in the likelihood surface linking it to the regime variance. The filtered regime probabilities are accurately recovered under correct specification, whereas the one-step-ahead conditional mean is insensitive to misspecification of the transition dynamics. An empirical application to U.S.\ Treasury zero-coupon yield changes (1961--2024) at four maturities shows that a specification driven by the lagged yield level provides the best fit, attaining the lowest AIC at all four maturities and the lowest BIC at the short end of the yield curve, and that the estimated regimes align with documented episodes of U.S.\ monetary history.

\medskip
\keywords{Markov-switching model, Time-varying transition probabilities, Generalized autoregressive score (GAS), Identifiability, Monte Carlo simulation, U.S.\ Treasury yields}
\msccodes{62M05 (primary), 62M10, 62F10, 91B84}
\jelcodes{C22, C15, C58, E43}
\end{abstract}

\newpage
\section{Introduction}\label{intro4}
\noindent

Macroeconomic and financial time series frequently exhibit nonlinear dynamics, structural change, and shifts in regimes or volatility.  To model nonlinear and regime-dependent dynamics in macroeconomic and financial time series, \citet{hamilton1989new} introduced the Markov Switching framework, in which occasional discrete shifts in an unobserved latent state govern changes in the behavior of the observed series. In its standard form, latent regimes evolve according to a first-order Markov chain with constant transition probabilities. This model has since then gained popularity and has been widely applied in macroeconomics and finance,  including detecting recessions, identifying different levels of volatility and classifying the stock market regime
\citep{Cai1994A,ravn1995stylized,chauvet1998econometric,garcia1996analysis,Bai2011Conditional,Guidolin2011Markov,Doornik2013A,Berentsen2022Modelling}. Subsequent research argued that the assumption of fixed transition probabilities is often too restrictive in empirical applications, because regime changes can depend on evolving economic conditions and other observable variables \citep{diebold1994regime,filardo1994business,Filardo1998Choosing}. \citet{diebold1994regime}
extended the \citet{hamilton1989new}
model by allowing transition probabilities to vary over time as functions of observable covariates, making the regime change responsive to economic behavior.
\citet{filardo1994business}
and
\citet{Filardo1998Choosing} further embedded time‑varying transition probabilities (TVTP) in the nonlinear Hamilton model, using observed “information variables” (economic or financial covariates) to drive the probability of moving between expansion and contraction in business cycles. \citet{filardo1994business} also showed that the Hidden Markov Model with TVTP can  track business cycles   more closely than constant probability models, and  macro variables can help to predict the turning point. The TVTP extension is thus attractive because time variation in transition probabilities can improve turning-point detection, the probabilistic assessment of regime changes, and economic interpretability.

In the TVTP Markov Switching (MS) models  proposed by  \citet{diebold1994regime} and \citet{filardo1994business}, the time-varying transition matrix is typically driven by exogenous variables.
\citet{creal2013generalized} introduced a general observation-driven framework in which time-varying parameters are updated by the scaled score of the predictive likelihood, providing a unified approach for a broad class of nonlinear models.   \citet{bazzi2017time}  adopted the score-driven framework of \citet{creal2013generalized} and established conditions under which the estimated time-varying probabilities converge to stationary and ergodic processes. Their empirical implementation is carried out on US industrial production growth data. The models in \citet{creal2013generalized} and \citet{bazzi2017time} belong to the Generalized Autoregressive Score (GAS) family of models, while the driver of the time‑varying transition matrix is the score of the conditional log‑likelihood. Based on how the  transition probabilities and Markov switching process are modeled, other types of TVTP-MS models are proposed, such as the time-varying and second-order Markov models proposed by \citet{Neale2016Regime} and the latent endogenous driver regime-switching model of \citet{chang2017new}, where the driver of the transition probability is a latent autoregressive factor. As TVTP-MS preserves Hamilton’s regime‑switching structure while allowing regime persistence and switching risks to respond to economic information or the previous data-stream, they can therefore further improve turning-point detection and model interpretability. Thus, TVTP-MS models are widely utilized, e.g., to identify the  recession/expansion in  business cycles \citep{diebold1994regime}; to accelerate the real-time dating of business-cycle peaks in a dynamic factor model whose transition probabilities are driven by the term spread \citep{vanos2024accelerating}; to model the volatility of crude-oil futures prices using basis-driven transition probabilities \citep{Fong2002A}; to model the conditional distribution of the short-term interest rate with switching probabilities driven by the lagged level of the rate \citep{gray1996modeling};
and to produce probabilistic forecasts of industrial electricity demand using time-varying transition probabilities driven by calendar variables \citep{Berk2018Probabilistic}. 

Among different TVTP-MS models, the GAS models studied in \citet{bazzi2017time}   are intuitive and economically interpretable.  The model parameters evolve over time in response to new information, and that information is measured by the score; thus the transition probabilities can adjust directly to the observations, and the updating is proportional to the information content in the recent data.
This model  has also demonstrated strong performance in Monte Carlo simulations and outperforms constant probability specifications in likelihood, information criteria, and forecasting. Although the implementation in \citet{bazzi2017time} is carried out to reveal the three regimes of industrial production growth data with three means and two variances, their simulation is carried out by Monte Carlo study for the two-regime models with two different means and common variance. In empirical macroeconomics a three-regime description is common, with the underlying states corresponding to recession, stability, and growth, each with its own mean and variance. It is therefore necessary to investigate how well the time-varying transition probability models can be estimated with more than two regimes with regime-specific means and variances,  and whether the filtered regime probabilities and transition dynamics can be recovered reliably in richer specifications.

The present paper addresses this gap by studying multi-regime Markov-switching models with TVTP under three alternative specifications of the transition matrix. It combines Monte Carlo evidence with an application to U.S. Treasury zero-coupon yield changes, a setting in which regime changes in volatility are of direct interest across maturities.  In our simulations, the transition probabilities follow three types of dynamics:
(a) the transition probabilities change with the lagged value of the observations;
(b) the transition probabilities change with an exogenous explanatory process;
(c) the transition probabilities evolve based on the score of the predictive likelihood.
These three specifications correspond to models (I), (II) and (III) in the remainder of this paper. We develop an open source R package \texttt{multiregimeTVTP} for the implementation of simulation in the general case of $K, K\geq 2$ regimes.  Different performance metrics are used to evaluate estimation performance, and a misspecification analysis is carried out to investigate the robustness of the fitted models to the choice of TVTP specification. The empirical implementation is carried out on  U.S.\ Treasury zero-coupon yield data constructed by \citet{liu2021reconstructing}. Thus, this paper has three main contributions. First, it extends the analysis of TVTP models to the general $K$-regime case with regime-specific means and variances and provides a unified open-source implementation for simulation and estimation. Second, it evaluates identification, parameter recovery, filtered probability accuracy, and robustness to misspecification in a systematic Monte Carlo design. Third, it shows in U.S. Treasury yield data that a level-driven TVTP specification provides the best in-sample fit among the models considered, and that the estimated regimes align with documented episodes of U.S.\ monetary history across maturities.

The remainder of this paper is divided into the following sections: Section 
\ref{sec:three models} introduces the general structure of the regime-switching model with  time-varying transition probabilities (TVTP) and the description of three types of TVTP models where the transition probabilities follow different dynamics. Section \ref{sec:sim} presents the estimation result based on comprehensive simulation studies for $K=2,3$ regimes, where each regime has its own mean and variance. Section \ref{sec:caseStudy} is devoted to the empirical analysis and the economic interpretation of our findings for the U.S.\ Treasury zero-coupon yield data. Section \ref{sec:conlusions} contains final conclusions and suggestions for future research.

\section{Regime-switching model with  time-varying transition probabilities}
\label{sec:three models}

\noindent 
We first recall the standard setting of regime-switching models with transition probabilities that do not vary with time. 
As in  \citet{bazzi2017time}, let $(y_1,y_2, \dots, y_t )$ denote a discrete-time time series of $t$ univariate observations, considered as a subset of a stochastic process $(y_t)$. 
Further, for $K \in \mathbb{N}$, consider a discrete Markov process $(z_t)$ with finite state space $\{1,2,\dots,K\}$ and  transition  matrix $\pi$, i.e.\ satisfying for each $i \in \{1,\dots,K\}$
 \begin{equation}
 \pi_{ij} =  P(z_{t}=j \mid z_{t-1}=i), j = 1,2,...,K, \text{ and } \sum_{j=1}^{K} \pi_{ij}=1,
 \end{equation}
 and where $\pi$ is assumed to be irreducible and aperiodic. This latter assumption implies that the fundamental theorem of Markov chains holds and hence $(z_t)$ is ergodic. 
 
In our setting, the Markov chain $(z_t)$ is a hidden process whose value is only observed indirectly by how it aﬀects the realizations of $(y_t)$. 
In particular, we assume that the random variables $(y_1,\dots,y_t)$ are conditionally independent given $(z_1,\dots,z_t)$ and with conditional distribution given by
\begin{equation} (y_{t}|z_{t}=i)\sim \mathcal{N}\left(\mu_{i},\sigma_{i}^{2}\right), \quad i =1,\dots, K.\end{equation}
Moreover, given the observed information available prior to time $t$, denoted by $I_{t-1} = \left\{y_{t-1}, y_{t-2},\dots \right\}$, and letting $\theta_i = \left(\mu_{i},\sigma_{i}^{2}\right)$ be the regime-specific parameters, the conditional distribution of $y_t$ can be written as
\begin{equation}
\mathbb{P}\left(y_t \in \cdot \mid I_{t-1}\right) = \sum^{K}_{i = 1}\sum^{K}_{k = 1}\mathbb{P}(y_t \in\cdot  \mid \theta_i)\pi_{ki}\mathbb{P}(z_{t-1} = k \mid I_{t-1}).
\end{equation}
This setting falls within the general theory of Markov regime-switching models, as presented in \citet{hamilton1994time}.
In particular, the Hamilton filter can be applied recursively to evaluate the likelihood, and the MLE is commonly used to estimate the parameters $\theta=(\theta_i,i=1,\dots,K)$. For details on the estimation procedure, we refer to \citet{hamilton1994time}, \citet{kim1999state} and \citet{bazzi2017time}. Consistency and asymptotic normality of the MLE in regime-switching models with time-varying transition probabilities have been established by \citet{Li2023Asymptotic}.

To extend the above setting to allow for the transition probabilities that vary with time, we consider transition probabilities for the (time-inhomogeneous) Markov chain $(z_t)$ that are driven, element by element, by real-valued dynamic processes $(f_{ij,t})$. In our main parameterization, each off-diagonal element of row $i$ of the transition matrix is obtained from its own scalar process $f_{ij,t}$ via the logistic transformation,
\begin{equation}\label{eq:logistic_link}
\pi_{ij,t}= \frac{1}{1+\exp(-{f_{ij,t}})}, \quad j \neq i, \qquad \text{and} \qquad \pi_{ii,t} = 1-\sum_{j \neq i}\pi_{ij,t}, \quad t\geq 1,
\end{equation}
so that each modeled probability lies strictly between $0$ and $1$, while the diagonal element is determined by the unit row-sum constraint. This yields $K(K-1)$ free transition parameters per time step, each of which may depend on the observed information prior to time $t$. We collect the modeled elements in $f_t = (f_{ij,t})_{i \neq j}$, and write $\pi_t = \pi(f_t)$ for the resulting transition matrix.
The case $t=1$ is considered the initialization of the process and the corresponding time-homogeneous Markov chain with transition matrix $\pi=\pi(f_1)$ is what we later refer to as the constant-case. 
 Further, in this paper we focus on the following  three concrete models to determine the dynamic processes $(f_{ij,t}; \: i \neq j, \: t\geq 1)$:
\begin{eqnarray}\label{Model1}
& {\rm Model\  (I):} & f_{ij,t}=\omega_{ij}+A_{ij}\, y_{t-1}
\\
& {\rm Model\  (II):} & f_{ij,t}=\omega_{ij}+ A_{ij}\, X_{t-1}\label{Model2}  \\
& {\rm Model\  (III):} &
f_{ij,t}=\omega_{ij}+A_{ij}\,s_{ij,t-1}+B_{ij}\,(f_{ij,t-1}-\omega_{ij}); \label{Model3}  \nonumber\\
&& s_{ij,t-1} =  S_{ij,t-1}\nabla_{ij,t-1}, \quad \nabla_{ij,t-1}= \frac{\partial}{\partial f_{ij,t-1}}\log p(y_{t-1} \mid I_{t-2};f_{t-1},\theta)
\end{eqnarray}

In model (I), $f_{ij,t}$ is simply a linear function of the lagged observation $y_{t-1}$, i.e.\ autoregressive dynamics where transitions depend on lagged observations. Written out element by element, each modeled transition element has its own scalar intercept $\omega_{ij}$ and slope coefficient $A_{ij}$, so that the scalar observation $y_{t-1}$ updates every free element of the transition matrix through its own pair of coefficients. Collecting these, $\omega = (\omega_{ij})$ and $A = (A_{ij})$ each contain $K(K-1)$ free parameters. The model extends  Hamilton’s Markov‑switching model by allowing transition probabilities to vary with the lagged dependent variable \citep{diebold1994regime,filardo1994business,bazzi2017time}.

Model (II) is a minor adaptation of model (I) where $f_{ij,t}$ is a linear function of the lagged value of an exogenous explanatory process $(X_t)$,
i.e.\ the transitions are driven by an external variable whose values are not influenced by the $(y_t)$ process. As in model (I), the intercepts $\omega_{ij}$ and coefficients $A_{ij}$ are element-specific scalars, giving $K(K-1)$ free parameters each. This model was proposed by \citet{diebold1994regime}, who performed a simulation study for the case of $K=2$ regimes, using the EM algorithm for parameter estimation.
 
 Model (III), where $f_{ij,t}$ is a linear function of the scaled score $s_{ij,t-1}$ and the previous value $f_{ij,t-1}$, was proposed by \citet{bazzi2017time} within the  Generalized Autoregressive Score (GAS) framework of \citet{creal2013generalized}. More specifically, in this score-driven dynamics, $s_{t} = (s_{ij,t})_{i\neq j}$ is the scaled score of the conditional observation density, i.e.\ the predictive likelihood, with respect to $f_{t}$. Each modeled element $f_{ij,t}$ is updated only by its own score $s_{ij,t-1}$ and its own lagged value, with element-specific coefficients $\omega_{ij}$, $A_{ij}$ and $B_{ij}$; stacking the modeled elements of $f_t$ into a vector, this corresponds to \emph{diagonal} loading matrices $A$ and $B$ in the notation of \citet{bazzi2017time}.
 
 Note that, by setting the scaling factors $S_{ij,t}$ equal to the square root of the inverse Fisher information, the scaled score $s_{t}$ has unit variance. As in models (I) and (II), $\omega_{ij}$ is the baseline level of $f_{ij,t}$---here its unconditional mean---and the specification is algebraically equivalent to the form $f_{ij,t} = w_{ij} + A_{ij}\, s_{ij,t-1} + B_{ij}\, f_{ij,t-1}$ used by \citet{bazzi2017time}, with $w_{ij} = \omega_{ij}(1-B_{ij})$; see Section~\ref{sec:sim_software} for implementation details. We refer to \citet{bazzi2017time}, Section 4, for further details on the statistical properties of the estimated dynamic transition probabilities of model (III). 
 
 \citet{creal2013generalized} pointed out that for model (III) the score  depends on the density function of a dataset, and it defines the steepest ascent direction for improving the model's local fit in terms of the likelihood at time $t$. Thus, it is intuitive to use the score to update $f_{{t}}$. In other words, model (III) can incorporate the information embedded in the conditional observation densities to the dynamics of transition probability. \citet{bazzi2017time} carried out a comprehensive simulation study for the case of $K=2$ regimes and showed that this model can indeed adequately track the dynamic patterns in the transition probabilities. 

The above three models of $f_{{t}}$ use informative observations of either $(y_t)$ or certain exogenous explanatory variables to determine the time varying  transition probabilities of $(z_t)$. In this paper we explore all three models of $f_{{t}}$ for the cases $K=2,3$  to capture the dynamics of the transition probabilities in a Monte Carlo based simulation study (Section \ref{sec:sim}) and an empirical study of U.S.\ Treasury yields (Section \ref{sec:caseStudy}). For completeness we also compare these with the constant, time-homogeneous, case.

\section{Monte Carlo simulation study}\label{sec:sim}

\noindent To validate the estimation procedure and assess parameter recovery performance under the three TVTP model specifications of Section~\ref{sec:three models}, we conduct a comprehensive Monte Carlo simulation study. The simulation covers the covariate-driven specifications of \citet{diebold1994regime} (Models~I and~II), where transition probabilities depend on lagged observables, and the score-driven framework of \citet{creal2013generalized} as implemented by \citet{bazzi2017time} (Model~III). While the Monte Carlo study in \citet{bazzi2017time} was limited to the two-regime case ($K=2$) with a common variance and different means for each regime, our study extends the analysis to three regimes ($K=3$) with regime-specific variances. In addition to correctly specified estimation, we systematically evaluate the estimation robustness to model misspecification by cross-estimating the three TVTP data-generating processes under all four model types (constant, Model~I, Model~II, and Model~III).

\subsection{Software implementation}\label{sec:sim_software}

\noindent We found no publicly available implementation covering Models~I--III jointly under the general $K$-regime, regime-specific-variance setting studied here, and therefore developed an open-source R package which provides a unified implementation of all four specifications together with the Monte Carlo infrastructure used to produce the results of this section. The package, which is called \texttt{multiregimeTVTP}, is available at \url{https://github.com/smodee/multiregime-TVTP}. The package was developed with the assistance of Claude (Anthropic, 2025--2026) for writing and refactoring R and C code; all code was reviewed, tested against the pure-R reference implementations and independent numerical checks, and validated by the authors, who take full responsibility for its correctness.

The package exposes a common interface for the four model families of Section~\ref{sec:three models}, with each model having matched \texttt{data<Model>CD()} simulators, \texttt{Rfiltering\_<Model>()} filters returning the log-likelihood and filtered probabilities, and \texttt{estimate\_<model>\_model()} estimators that wrap \texttt{optim} with multi-start initialization. All specifications support arbitrary $K \geq 2$ and both the diagonal and off-diagonal parameterizations of the transition matrix. In the off-diagonal parameterization, each modeled element is transformed independently via the logistic link of equation~(\ref{eq:logistic_link}); as a numerical safeguard, if the off-diagonal probabilities of a row would sum to more than one, they are proportionally rescaled so that the transition matrix remains valid. Wald standard errors on the original scale are recovered from the numerical Hessian (\texttt{numDeriv}) via the delta method. 

For the Monte Carlo study, higher-level analysis functions coordinate the full data-generating process $\times$ sample-size $\times$ estimation-model grid, the $R$ replications, and the $n$ random starts per fit. Replication-level estimates are aggregated into the bias, RMSE, coverage, one-step prediction error, and filtered-probability summaries of Section~\ref{sec:sim_metrics}. Multi-start optimization is parallelized through the \texttt{future}/\texttt{future.apply} backends, so that replications and starts run concurrently across cores. The four filtering routines, which are called at every likelihood evaluation, are additionally backed by a compiled C implementation (\texttt{src/filtering.c}) that yields a 10--56$\times$ speedup per likelihood call relative to the pure-R filter, with the largest gains on Models~I and~III. The backend is selected automatically when available and can be toggled for benchmarking, which made the full Monte Carlo study tractable (days of compute time instead of weeks/months) on standard desktop hardware while leaving the pure-R filters available as a reference implementation. The same estimation, filtering, and analysis routines are reused without modification in the empirical application of Section~\ref{sec:caseStudy}.

Model~III is implemented directly in the mean-reverting form of equation~(\ref{Model3}). The filter is initialized at $f_1 = \omega$, so $A \to 0$ collapses $f_t$ to the constant $\omega$ and the model reduces exactly to constant transition probabilities $\pi_{ij} = \text{logistic}(\omega)$. We exploit this by falling back to the constant filter whenever $\max_{ij}|A_{ij}|$ drops below a small numerical threshold, which also avoids near-flat likelihood regions during optimization.

\subsection{Simulation design}\label{sec:sim_design}

We consider nine data-generating processes (DGPs), summarised in Table~\ref{tab:mc_scenarios}, each evaluated at two sample sizes $T\in\{500,\, 1{,}000\}$, giving 18 scenario--sample-size combinations in total. The DGPs are arranged in two parallel blocks: a $K=2$ block (DGPs~1--4) with diagonal transition parameterization and common variance, and a $K=3$ block (DGPs~6--9) with off-diagonal parameterization and regime-specific variances. Within each block, the first DGP (1 and~6) is a constant-transition baseline, estimated only under the constant specification; the remaining three (2--4 and~7--9) are generated by TVTP Models~I, II, and~III respectively and estimated under all four model types, yielding the cross-TVTP misspecification analysis. Fitting Models~I--III to the constant baselines would only test whether the dynamic coefficients shrink to zero, which is a separate question from the cross-TVTP comparison targeted here. DGP~5 is an additional $K=2$ scenario that keeps Model~I as the data-generating process but adopts the off-diagonal, regime-specific-variance parameterization of the $K=3$ block. It is estimated only under the constant and Model~I specifications, to isolate the effect of the parameterization change rather than repeat the misspecification comparison already covered by DGPs~2--4.

\begin{table}[!htbp]
\centering
\caption{Monte Carlo data-generating processes with each DGP evaluated at $T\in\{500,\, 1{,}000\}$. ``Diag'' indicates diagonal transition parameterization; ``Eq.\ var'' indicates common variance across regimes. }
\label{tab:mc_scenarios}
\small
\begin{tabular}{clcccl}
\toprule
DGP & Model & $K$ & Diag & Eq.\ var & Estimation models \\
\midrule
1 & Constant   & 2 & \checkmark & \checkmark & Const \\
2 & Model~I    & 2 & \checkmark & \checkmark & Const, I, II, III \\
3 & Model~II   & 2 & \checkmark & \checkmark & Const, I, II, III \\
4 & Model~III  & 2 & \checkmark & \checkmark & Const, I, II, III \\
5 & Model~I    & 2 &            &            & Const, I \\
\midrule
6 & Constant   & 3 &            &            & Const \\
7 & Model~I    & 3 &            &            & Const, I, II, III \\
8 & Model~II   & 3 &            &            & Const, I, II, III \\
9 & Model~III  & 3 &            &            & Const, I, II, III \\
\bottomrule
\end{tabular}
\end{table}

The true parameter values used to generate the data are reported in Table~\ref{tab:true_params}. Within the $K=2$ block, DGPs~1--4 share regime means $\mu = (-1, 1)$ and a common variance $\sigma^2 = 0.5$; DGP~5 retains the same means but uses regime-specific variances $\sigma^2 = (0.3, 0.7)$. Within the $K=3$ block (DGPs~6--9), all DGPs share regime means $\mu = (-2, 0, 2)$ and regime-specific variances $\sigma^2 = (0.3, 0.5, 0.8)$. The TVTP coefficient magnitudes are moderate, representing empirically plausible effect sizes that allow the transition probabilities to vary meaningfully over time without dominating the regime dynamics.

\begin{table}[!htbp]
\centering
\caption{True parameter values per DGP. Diagonal parameterization for DGPs~1--4; off-diagonal ($K(K-1)$ transition parameters) for DGPs~5--9. Intercept column: baseline transition probabilities $\pi^{(0)}_{ij}$, which determine the intercepts $\omega_{ij}$ of equations~(\ref{Model1})--(\ref{Model3}); for $A=0$, $\pi_{ij,t}=\pi^{(0)}_{ij}$. Columns $A$ and $B$: dynamic TVTP coefficients in the notation of Section~\ref{sec:three models}; $B$ applies to Model~III only.}
\label{tab:true_params}
\footnotesize
\begin{tabular}{cl l l l}
\toprule
DGP & Model & Intercept & $A$ & $B$ \\
\midrule
1 & Constant  & $\pi^{(0)}_{11}=0.80$,\; $\pi^{(0)}_{22}=0.90$ & & \\
2 & Model~I   & $\pi^{(0)}=(0.80,\, 0.90)$ & $(0.15,\, {-}0.10)$ & \\
3 & Model~II  & $\pi^{(0)}=(0.80,\, 0.90)$ & $(0.20,\, {-}0.20)$ & \\
4 & Model~III & $\pi^{(0)}=(0.80,\, 0.90)$ & $(0.10,\, {-}0.10)$ & $(0.90,\, 0.85)$ \\
5 & Model~I   & $\pi^{(0)}_{\text{off}}=(0.20,\, 0.15)$ & $(0.10,\, {-}0.10)$ & \\
\midrule
6 & Constant  & \multicolumn{3}{l}{$\pi^{(0)}_{\text{off}} = (0.08,\, 0.08,\, 0.10,\, 0.10,\, 0.06,\, 0.06)$} \\
7 & Model~I   & $\pi^{(0)}_{\text{off}}$ (as DGP~6) & $(0.05,\, {-}0.03,\, 0.04,\, {-}0.04,\, 0.03,\, {-}0.05)$ & \\
8 & Model~II  & $\pi^{(0)}_{\text{off}}$ (as DGP~6) & $(0.08,\, {-}0.04,\, 0.05,\, {-}0.06,\, 0.04,\, {-}0.07)$ & \\
9 & Model~III & $\pi^{(0)}_{\text{off}}$ (as DGP~6) & all elements $0.03$ & all elements $0.85$ \\
\bottomrule
\end{tabular}
\end{table}

For DGPs with TVTP dynamics, each dataset is estimated not only under the correctly specified model but also under the remaining model types. This misspecification design, analogous to that of \citet{bazzi2017time}, allows us to assess how Models~I--III perform when the true dynamics follow a different specification. For Model~II estimation on data not generated by an exogenous process, we supply an independent standard normal series as the exogenous variable, representing the case where the covariate carries no information about the true regime dynamics.

Each DGP--sample-size combination is replicated $R=50$ times. For every replication, the model is estimated by maximum likelihood using $n=10$ random starting points, keeping the best result (by log-likelihood among converged runs). In every filtering pass, the first 100 observations serve as a burn-in: they are processed by the filter, so that the influence of its initialization dies out, but are excluded from the likelihood sum. The last 10 observations are likewise excluded as an end-of-sample cut-off. At $T=500$, for example, the likelihood is thus accumulated over the remaining 390 observations. To address the label-switching problem inherent in mixture and regime-switching models, we align estimated regimes to the true ordering by finding the permutation of regime labels that minimizes the sum of absolute deviations between estimated and true regime means.

\subsection{Performance metrics}\label{sec:sim_metrics}

We evaluate estimation performance along four dimensions, following and extending the metrics reported in \citet{bazzi2017time}.

\paragraph{Parameter recovery.} For each model parameter $\vartheta \in \{\mu_i, \sigma_i^2, \pi_{ij}, A_{ij}, \ldots\}$, with true value $\vartheta_0$, we compute the bias $\text{Bias}(\hat\vartheta) = R^{-1}\sum_{r=1}^{R}(\hat\vartheta_r - \vartheta_0)$ and root mean squared error $\text{RMSE}(\hat\vartheta) = \bigl[R^{-1}\sum_{r=1}^{R}(\hat\vartheta_r - \vartheta_0)^2\bigr]^{1/2}$ across replications.

\paragraph{Coverage rates.} Standard errors are obtained from the numerical Hessian of the log-likelihood (evaluated at the MLE in the transformed parameter space) via the \texttt{numDeriv} package, with the delta method applied for parameters subject to log or logit transformations. We report the empirical coverage rate of nominal 95\% Wald confidence intervals.

\paragraph{One-step-ahead prediction error.} Following \citet{bazzi2017time}, we compute the one-step-ahead conditional means $\hat{y}_{t|t-1} = \sum_{i=1}^{K} \hat\mu_i \, \hat{P}(z_t = i \mid I_{t-1})$ and evaluate them using the mean absolute forecast error (MAFE) and the mean squared forecast error (MSFE). This metric serves as a diagnostic of how strongly the fitted conditional mean depends on the specification of the transition dynamics.

\paragraph{Filtered probability accuracy.} For correctly specified estimations, we obtain reference filtered regime probabilities $P^{*}(z_t = k \mid I_t)$ by running the filter at the true parameter values, and compute the mean squared error and mean absolute error of the estimated filtered regime probabilities $\hat{P}(z_t = k \mid I_t)$ relative to this reference, averaged over regimes and time periods (after the regime alignment described in Section~\ref{sec:sim_design}).

\subsection{Simulation results}\label{sec:sim_results}

\noindent We first present parameter recovery and coverage results for correctly specified models, then compare the one-step-ahead prediction errors across estimation models to assess robustness to misspecification.

\subsubsection*{Parameter recovery}

Table~\ref{tab:param_recovery} reports average bias and RMSE for each parameter group under correct specification, with the true parameter values given in Table~\ref{tab:true_params}. For DGPs~1 and~6 (constant transition probabilities), the regime means $\mu_i$, variances $\sigma_i^2$, and transition probabilities $\pi_{ij}$ are all recovered with small bias and RMSE that decrease as the sample size doubles from $T=500$ to $T=1{,}000$, confirming consistency. For Model~I (DGPs~2, 5, 7), the distribution parameters $\mu_i$ and $\sigma_i^2$ are similarly well recovered, but the TVTP coefficient vector $A$ exhibits substantially larger RMSE, particularly for $K=3$ regimes. At $T=500$, 4 of 44 convergent replications in DGP~7 exhibit extreme $A$ estimates ($|\hat{A}_{ij}|>10$, against true values of order $0.05$), indicating optimizer breakdown on a flat likelihood region rather than genuine poor estimation. These replications are excluded from the reported $A$ statistics; after exclusion the trimmed RMSE is $1.550$, and increasing the sample to $T=1{,}000$ reduces it further to $0.549$. The off-diagonal parameterization of DGP~5 also yields elevated $A$ RMSE at $T=500$ ($1.228$), which improves to $0.319$ at $T=1{,}000$.

For Model~II (DGPs~3 and~8), parameter recovery shows a similar pattern: the $A$ coefficients have larger RMSE than the distribution parameters, with clear improvement at $T=1{,}000$. In the two-regime setting (DGP~3), the $A$ RMSE drops from $0.311$ to $0.189$; in the three-regime setting (DGP~8), it drops from $0.815$ to $0.298$.

For Model~III (DGPs~4 and~9), the distribution parameters $\mu_i$ and $\sigma_i^2$ and the transition probabilities $\pi_{ij}$ are well recovered, with bias and RMSE comparable to the other TVTP models. However, the score coefficient matrix $A$ is not reliably identified. The RMSE equals the magnitude of the true $A$ values and shows no improvement from $T=500$ to $T=1{,}000$ in either DGP, pointing to a statistical identifiability problem inherent to the GAS specification rather than a computational artefact. Profile likelihood analysis confirms this, showing that the 1D NLL is lower at $A=0$ than at the true value and that the MLE is at $A\approx 0$ regardless of the data-generating parameter. Examining the joint $(\sigma^2, A)$ space through a 2D profile further reveals a pronounced ridge in the likelihood surface, arising because the GAS score scaling couples $\sigma^2$ (through the Fisher information) and $A$ (as a direct multiplier) such that many parameter combinations yield nearly identical filtered transition probabilities and log-likelihoods.

\begin{table}[!htbp]
\centering
\caption{Parameter recovery for correctly specified models. Values are averages across parameter elements within each group. $R_c$ denotes convergent replications out of 50.}
\label{tab:param_recovery}
\footnotesize
\begin{tabular}{cl c r r rr rr rr rr}
\toprule
& & & & & \multicolumn{2}{c}{$\mu$} & \multicolumn{2}{c}{$\sigma^2$} & \multicolumn{2}{c}{$\pi$} & \multicolumn{2}{c}{$A$} \\
\cmidrule(lr){6-7} \cmidrule(lr){8-9} \cmidrule(lr){10-11} \cmidrule(lr){12-13}
DGP & Model & $K$ & $T$ & $R_c$ & Bias & RMSE & Bias & RMSE & Bias & RMSE & Bias & RMSE \\
\midrule
\multirow{2}{*}{1} & \multirow{2}{*}{Const.} & \multirow{2}{*}{2}
  & 500  & 50 & 0.016 & 0.057 & 0.001 & 0.044 & $-$0.008 & 0.030 & & \\
  & & & 1000 & 50 & 0.000 & 0.040 & 0.001 & 0.026 & 0.002 & 0.022 & & \\
\midrule
\multirow{2}{*}{2} & \multirow{2}{*}{I} & \multirow{2}{*}{2}
  & 500  & 49 & 0.000 & 0.060 & $-$0.007 & 0.043 & $-$0.013 & 0.096 & $-$0.032 & 0.423 \\
  & & & 1000 & 50 & 0.001 & 0.044 & $-$0.001 & 0.029 & 0.025 & 0.080 & 0.065 & 0.289 \\
\midrule
\multirow{2}{*}{3} & \multirow{2}{*}{II} & \multirow{2}{*}{2}
  & 500  & 50 & 0.005 & 0.069 & 0.002 & 0.036 & 0.000 & 0.074 & $-$0.010 & 0.311 \\
  & & & 1000 & 50 & 0.003 & 0.042 & 0.007 & 0.026 & $-$0.011 & 0.048 & $-$0.032 & 0.189 \\
\midrule
\multirow{2}{*}{4} & \multirow{2}{*}{III} & \multirow{2}{*}{2}
  & 500  & 50 & $-$0.003 & 0.058 & 0.010 & 0.040 & 0.002 & 0.037 & 0.000 & 0.100$^{\ddagger}$ \\
  & & & 1000 & 49 & $-$0.007 & 0.039 & 0.003 & 0.029 & 0.005 & 0.027 & 0.000 & 0.100$^{\ddagger}$ \\
\midrule
\multirow{2}{*}{5} & \multirow{2}{*}{I} & \multirow{2}{*}{2}
  & 500  & 48 & $-$0.008 & 0.071 & $-$0.003 & 0.071 & 0.022 & 0.118 & $-$0.083 & 1.228 \\
  & & & 1000 & 44 & 0.001 & 0.039 & $-$0.001 & 0.042 & 0.005 & 0.099 & 0.036 & 0.319 \\
\midrule
\multirow{2}{*}{6} & \multirow{2}{*}{Const.} & \multirow{2}{*}{3}
  & 500  & 50 & 0.000 & 0.086 & 0.004 & 0.104 & 0.001 & 0.032 & & \\
  & & & 1000 & 50 & $-$0.001 & 0.045 & $-$0.007 & 0.052 & 0.003 & 0.021 & & \\
\midrule
\multirow{2}{*}{7} & \multirow{2}{*}{I} & \multirow{2}{*}{3}
  & 500  & 44 & $-$0.011 & 0.106 & 0.052 & 0.177 & 0.060 & 0.165 & $-$0.320$^{\dagger}$ & 1.550$^{\dagger}$ \\
  & & & 1000 & 50 & $-$0.010 & 0.053 & 0.004 & 0.071 & 0.022 & 0.114 & $-$0.011 & 0.549 \\
\midrule
\multirow{2}{*}{8} & \multirow{2}{*}{II} & \multirow{2}{*}{3}
  & 500  & 50 & $-$0.003 & 0.079 & $-$0.013 & 0.085 & 0.005 & 0.066 & $-$0.076 & 0.815 \\
  & & & 1000 & 50 & 0.000 & 0.052 & $-$0.004 & 0.056 & 0.003 & 0.048 & $-$0.013 & 0.298 \\
\midrule
\multirow{2}{*}{9} & \multirow{2}{*}{III} & \multirow{2}{*}{3}
  & 500  & 50 & 0.004 & 0.088 & $-$0.005 & 0.088 & 0.006 & 0.034 & $-$0.030 & 0.030$^{\ddagger}$ \\
  & & & 1000 & 50 & 0.003 & 0.050 & $-$0.011 & 0.056 & 0.002 & 0.019 & $-$0.030 & 0.030$^{\ddagger}$ \\
\bottomrule
\multicolumn{13}{l}{\parbox{0.98\textwidth}{\scriptsize $\dagger$ 4 of 44 convergent reps.\ excluded due to optimizer breakdown ($|\hat{A}_{ij}|>10$); $A$ statistics based on 40 retained reps.}} \\
\multicolumn{13}{l}{\parbox{0.98\textwidth}{\scriptsize $\ddagger$ $A$ RMSE equals the true parameter magnitude and does not improve with $T$; see text.}}
\end{tabular}
\end{table}

\subsubsection*{Coverage rates}

Table~\ref{tab:coverage} reports the average empirical coverage rates of nominal 95\% Wald confidence intervals. For the constant-probability DGPs (1 and~6), coverage rates for $\mu$ and $\sigma^2$ are close to the nominal level across both sample sizes, though $\sigma^2$ coverage in the three-regime case (DGP~6) drops to $0.867$ at $T=500$ before recovering to $0.960$ at $T=1{,}000$.

Under TVTP specifications, the coverage for transition probability parameters $\pi$ is systematically below the nominal 95\% level, particularly for $K=3$ models. In DGPs~7 and~8 at $T=500$, the average $\pi$ coverage is only $0.675$ and $0.713$, respectively. While coverage improves at $T=1{,}000$ ($0.753$ and $0.817$), it remains substantially below nominal. 
Coverage for $\mu$ remains robust across all settings, generally exceeding $0.90$. For Model~III (DGP~4), $\sigma^2$ coverage reaches $1.000$ at $T=500$, consistent with a numerically ill-conditioned Hessian along the $(\sigma^2, A)$ ridge identified in the parameter recovery discussion above: the near-flat likelihood inflates the computed standard errors, producing confidence intervals wide enough to contain the true value in virtually every replication.

\begin{table}[!htbp]
\centering
\caption{Average empirical coverage of nominal 95\% Wald confidence intervals for correctly specified models.}
\label{tab:coverage}
\small
\begin{tabular}{cl cc rrr}
\toprule
DGP & Model & $K$ & $T$ & $\mu$ & $\sigma^2$ & $\pi$ \\
\midrule
\multirow{2}{*}{1} & \multirow{2}{*}{Const.} & \multirow{2}{*}{2}
  & 500  & 0.960 & 0.960 & 0.960 \\
  & & & 1000 & 0.980 & 0.960 & 0.910 \\
\midrule
\multirow{2}{*}{2} & \multirow{2}{*}{I} & \multirow{2}{*}{2}
  & 500  & 0.980 & 0.898 & 0.837 \\
  & & & 1000 & 0.960 & 0.920 & 0.870 \\
\midrule
\multirow{2}{*}{3} & \multirow{2}{*}{II} & \multirow{2}{*}{2}
  & 500  & 0.920 & 0.980 & 0.870 \\
  & & & 1000 & 0.940 & 0.960 & 0.950 \\
\midrule
\multirow{2}{*}{4} & \multirow{2}{*}{III} & \multirow{2}{*}{2}
  & 500  & 0.960 & 1.000 & 0.890 \\
  & & & 1000 & 0.949 & 0.980 & 0.857 \\
\midrule
\multirow{2}{*}{5} & \multirow{2}{*}{I} & \multirow{2}{*}{2}
  & 500  & 0.917 & 0.927 & 0.844 \\
  & & & 1000 & 0.966 & 0.955 & 0.920 \\
\midrule
\multirow{2}{*}{6} & \multirow{2}{*}{Const.} & \multirow{2}{*}{3}
  & 500  & 0.933 & 0.867 & 0.877 \\
  & & & 1000 & 0.960 & 0.960 & 0.927 \\
\midrule
\multirow{2}{*}{7} & \multirow{2}{*}{I} & \multirow{2}{*}{3}
  & 500  & 0.908 & 0.850 & 0.675 \\
  & & & 1000 & 0.947 & 0.887 & 0.753 \\
\midrule
\multirow{2}{*}{8} & \multirow{2}{*}{II} & \multirow{2}{*}{3}
  & 500  & 0.933 & 0.907 & 0.713 \\
  & & & 1000 & 0.960 & 0.940 & 0.817 \\
\midrule
\multirow{2}{*}{9} & \multirow{2}{*}{III} & \multirow{2}{*}{3}
  & 500  & 0.940 & 0.940 & 0.927 \\
  & & & 1000 & 0.940 & 0.960 & 0.937 \\
\bottomrule
\end{tabular}
\end{table}

\subsubsection*{Robustness to misspecification}

Table~\ref{tab:forecast} compares the one-step-ahead prediction errors across estimation models for each DGP, under both correct specification and misspecification. Across all DGPs and sample sizes, the metrics are remarkably stable: MAFE and MSFE differ by less than 1\% whether the correctly specified or a misspecified model is fitted. This finding extends the result of \citet{bazzi2017time} from two regimes to three, and its explanation is the same. The conditional mean $\hat{y}_{t|t-1} = \sum_i \hat{\mu}_i \hat{P}(z_t = i \mid I_{t-1})$ is dominated by the regime means $\hat{\mu}_i$, which are robustly recovered under all specifications. Since the transition dynamics have only a marginal effect on the predicted regime probabilities over a single step, misspecifying them leaves the conditional mean essentially unchanged.

The one-step conditional mean is therefore an insensitive diagnostic of specification error, and its stability should not be read as implying that the choice of transition dynamics is inconsequential. The object that does respond to the specification is the filtered regime path: the results below show that the latent regimes are accurately reconstructed under correct specification, and the empirical application of Section~\ref{sec:caseStudy} shows that the transition specification has a substantial effect on model fit.

\begin{table}[!htbp]
\centering
\caption{One-step-ahead prediction error (MAFE and MSFE) across estimation models. Bold indicates the correctly specified model.}
\label{tab:forecast}
\footnotesize
\begin{tabular}{cl c rrrr rrrr}
\toprule
& & & \multicolumn{4}{c}{MAFE} & \multicolumn{4}{c}{MSFE} \\
\cmidrule(lr){4-7} \cmidrule(lr){8-11}
DGP & $T$ & $K$ & Const. & I & II & III & Const. & I & II & III \\
\midrule
1 & 500  & 2 & \textbf{0.795} & & & & \textbf{1.113} & & & \\
1 & 1000 & 2 & \textbf{0.786} & & & & \textbf{1.077} & & & \\
\midrule
2 & 500  & 2 & 0.783 & \textbf{0.784} & 0.783 & 0.783 & 1.085 & \textbf{1.086} & 1.085 & 1.085 \\
2 & 1000 & 2 & 0.782 & \textbf{0.782} & 0.782 & 0.782 & 1.079 & \textbf{1.079} & 1.079 & 1.079 \\
\midrule
3 & 500  & 2 & 0.748 & 0.748 & \textbf{0.748} & 0.748 & 0.981 & 0.982 & \textbf{0.980} & 0.981 \\
3 & 1000 & 2 & 0.751 & 0.750 & \textbf{0.750} & 0.751 & 0.989 & 0.989 & \textbf{0.988} & 0.989 \\
\midrule
4 & 500  & 2 & 0.806 & 0.806 & 0.806 & \textbf{0.806} & 1.130 & 1.130 & 1.130 & \textbf{1.130} \\
4 & 1000 & 2 & 0.798 & 0.798 & 0.798 & \textbf{0.797} & 1.115 & 1.115 & 1.115 & \textbf{1.112} \\
\midrule
5 & 500  & 2 & 0.837 & \textbf{0.839} & & & 1.213 & \textbf{1.218} & & \\
5 & 1000 & 2 & 0.837 & \textbf{0.838} & & & 1.215 & \textbf{1.217} & & \\
\midrule
6 & 500  & 3 & \textbf{0.952} & & & & \textbf{1.872} & & & \\
6 & 1000 & 3 & \textbf{0.961} & & & & \textbf{1.891} & & & \\
\midrule
7 & 500  & 3 & 0.954 & \textbf{0.963} & 0.954 & 0.954 & 1.897 & \textbf{1.920} & 1.889 & 1.897 \\
7 & 1000 & 3 & 0.970 & \textbf{0.970} & 0.969 & 0.970 & 1.966 & \textbf{1.966} & 1.965 & 1.966 \\
\midrule
8 & 500  & 3 & 0.957 & 0.958 & \textbf{0.957} & 0.957 & 1.887 & 1.892 & \textbf{1.885} & 1.887 \\
8 & 1000 & 3 & 0.963 & 0.963 & \textbf{0.963} & 0.963 & 1.913 & 1.914 & \textbf{1.912} & 1.913 \\
\midrule
9 & 500  & 3 & 0.967 & 0.969 & 0.967 & \textbf{0.967} & 1.923 & 1.925 & 1.923 & \textbf{1.923} \\
9 & 1000 & 3 & 0.960 & 0.962 & 0.960 & \textbf{0.960} & 1.888 & 1.890 & 1.887 & \textbf{1.888} \\
\bottomrule
\end{tabular}
\end{table}

\subsubsection*{Filtered probability accuracy}

Table~\ref{tab:filt_prob} reports the accuracy of the filtered regime probabilities for correctly specified models, measured by mean squared error (MSE) and mean absolute error (MAE) relative to the reference probabilities obtained by running the filter at the true parameter values. Accuracy is high throughout: the MSE stays below $0.008$ across all DGPs and sample sizes, and it decreases from $T=500$ to $T=1{,}000$ in every DGP, in line with the parameter recovery results of Table~\ref{tab:param_recovery}. The three-regime DGPs are somewhat harder than their two-regime counterparts---at $T=500$ the MSE ranges from $0.0023$ (DGP~9) to $0.0079$ (DGP~7), against roughly $0.001$ in the diagonal $K=2$ DGPs (1--4)---reflecting the larger number of transition parameters ($K(K-1)=6$) and the overlap between adjacent regime distributions, but the differences are of degree rather than kind. The largest errors occur for Model~I with $K=3$ (DGP~7), the DGP whose TVTP coefficients are also the hardest to recover. Filtered regime probabilities are thus reliably reconstructed whenever the specification is correct and the parameters are well estimated.

\begin{table}[!htbp]
\centering
\caption{Filtered probability accuracy for correctly specified models. MSE and MAE of the filtered regime probabilities, computed relative to the filter run at the true parameter values and averaged over regimes and time periods.}
\label{tab:filt_prob}
\small
\begin{tabular}{cl cc rr}
\toprule
DGP & Model & $K$ & $T$ & MSE & MAE \\
\midrule
\multirow{2}{*}{1} & \multirow{2}{*}{Const.} & \multirow{2}{*}{2}
  & 500  & 0.0008 & 0.0114 \\
  & & & 1000 & 0.0004 & 0.0083 \\
\midrule
\multirow{2}{*}{2} & \multirow{2}{*}{I} & \multirow{2}{*}{2}
  & 500  & 0.0012 & 0.0135 \\
  & & & 1000 & 0.0005 & 0.0088 \\
\midrule
\multirow{2}{*}{3} & \multirow{2}{*}{II} & \multirow{2}{*}{2}
  & 500  & 0.0012 & 0.0125 \\
  & & & 1000 & 0.0005 & 0.0083 \\
\midrule
\multirow{2}{*}{4} & \multirow{2}{*}{III} & \multirow{2}{*}{2}
  & 500  & 0.0010 & 0.0128 \\
  & & & 1000 & 0.0005 & 0.0089 \\
\midrule
\multirow{2}{*}{5} & \multirow{2}{*}{I} & \multirow{2}{*}{2}
  & 500  & 0.0039 & 0.0222 \\
  & & & 1000 & 0.0008 & 0.0113 \\
\midrule
\multirow{2}{*}{6} & \multirow{2}{*}{Const.} & \multirow{2}{*}{3}
  & 500  & 0.0029 & 0.0190 \\
  & & & 1000 & 0.0008 & 0.0100 \\
\midrule
\multirow{2}{*}{7} & \multirow{2}{*}{I} & \multirow{2}{*}{3}
  & 500  & 0.0079 & 0.0274 \\
  & & & 1000 & 0.0018 & 0.0141 \\
\midrule
\multirow{2}{*}{8} & \multirow{2}{*}{II} & \multirow{2}{*}{3}
  & 500  & 0.0032 & 0.0183 \\
  & & & 1000 & 0.0013 & 0.0121 \\
\midrule
\multirow{2}{*}{9} & \multirow{2}{*}{III} & \multirow{2}{*}{3}
  & 500  & 0.0023 & 0.0166 \\
  & & & 1000 & 0.0009 & 0.0101 \\
\bottomrule
\end{tabular}
\end{table}

\subsubsection*{Summary of Monte Carlo simulation}

The Monte Carlo results demonstrate that the estimation procedure reliably recovers the distribution parameters ($\mu_i$, $\sigma_i^2$) and transition probabilities across both $K=2$ and $K=3$ settings, with performance improving at larger sample sizes as expected. The TVTP driving coefficients ($A$ for Models~I and~II) are the most challenging parameters to estimate, requiring $T=1{,}000$ observations for the three-regime case to achieve acceptable RMSE levels. For Model~III (GAS), the score coefficient $A$ is statistically non-identifiable due to a ridge in the joint $(\sigma^2, A)$ likelihood surface; the remaining parameters are nevertheless well recovered. Coverage rates for transition probabilities under TVTP models are systematically below the nominal 95\%, particularly for $K=3$, indicating that alternative standard error procedures (e.g., bootstrap or sandwich estimators) may be warranted in practice.

The misspecification analysis shows that the one-step-ahead conditional mean is insensitive to the choice of TVTP specification, consistent with \citet{bazzi2017time}. This robustness is a mechanical consequence of the regime means dominating the conditional mean over a single step, and should not be read as implying that model choice is inconsequential. The filtered regime probabilities are accurately recovered under correct specification, with errors shrinking as the sample grows. The primary benefit of correct TVTP specification thus lies in accurately characterising the time-varying regime dynamics.

\section{Empirical analysis}\label{sec:caseStudy}

\subsection{Data}\label{sec:data}

\noindent We apply the three-regime TVTP models to U.S.\ Treasury zero-coupon yield data reconstructed by \citet{liu2021reconstructing}. The dataset contains monthly annualized continuously-compounded zero-coupon yields for maturities ranging from 1 to 360 months. Our sample spans June 1961 to December 2024, providing $T = 763$ monthly observations. We select four representative maturities along the yield curve: 1 month (short end), 12 months (short-to-medium), 36 months (medium), and 72 months (long end).

Following standard practice for yield curve modelling, we work with first differences of the yield series, i.e.\ $y_t = Y_t - Y_{t-1}$, where $Y_t$ denotes the yield level at time $t$. This transformation produces $T = 762$ observations of monthly yield changes for each maturity. First-differencing removes the strong persistence in yield levels \citep{hamilton1994time} and produces a more nearly stationary series suitable for regime-switching analysis.

\subsection{Model specification}\label{sec:empirical_spec}

\noindent For each maturity, we estimate a three-regime ($K=3$) Markov switching model where the conditional distribution of yield changes in regime $i$ is
\begin{equation}
(y_t \mid z_t = i) \sim \mathcal{N}(\mu_i, \sigma_i^2), \quad i = 1, 2, 3,
\end{equation}
with regime-specific means $\mu_i$ and variances $\sigma_i^2$. The three regimes are intended to capture distinct yield curve dynamics: a high-volatility regime associated with large yield movements, a moderate regime representing normal market conditions, and a low-volatility regime reflecting periods of relative stability.

We consider four model specifications for the transition probability dynamics:
\begin{itemize}
	\item \textbf{Constant:} The baseline model with time-invariant transition probabilities, corresponding to the classical Hamilton framework \citep{hamilton1989new} extended to three regimes.
	\item \textbf{Model (I) -- TVP:} Time-varying transition probabilities driven by the lagged observation $y_{t-1}$, as in equation (\ref{Model1}). This specification allows the most recent yield change to inform the probability of transitioning between regimes.
	\item \textbf{Model (II) -- Exogenous:} Time-varying transition probabilities driven by a covariate $X_{t-1}$, as in equation (\ref{Model2}). Here we set $X_{t-1} = Y_{t-1}$, the lagged yield level, so that regime dynamics are linked directly to the \emph{level} of interest rates rather than the yield \emph{change} used in the TVP specification. We note that the lagged level is predetermined rather than strictly exogenous, being a function of the history of the same series whose changes are modeled; ``exogenous'' here refers to the model class of equation (\ref{Model2}), and what distinguishes this specification from Model (I) is that the transition dynamics respond to the level rather than to the most recent change.
	\item \textbf{Model (III) -- GAS:} Generalized Autoregressive Score transition probabilities, as in equation (\ref{Model3}), where the transition parameters are updated using the scaled score of the conditional observation density.
\end{itemize}

The transition probabilities are parameterized via the logistic link function as described in Section~\ref{sec:three models}, with $K(K-1) = 6$ unconstrained parameters governing the off-diagonal elements of the transition matrix. Each model is estimated by maximum likelihood using $n = 100$ random starting points. As in the simulation study, the first 100 observations serve as a filter burn-in and are excluded from the likelihood; no end-of-sample cut-off is applied here. The log-likelihoods, AIC, and BIC reported below are therefore based on the remaining $662$ of the $T = 762$ observations. The best result across starting points (by log-likelihood among converged runs) is retained.

\subsection{Results}\label{sec:empirical_results}

\noindent We attempted to estimate the GAS specification (Model~III) alongside the other models. However, across all four maturities, the GAS model exhibited severe convergence difficulties: out of 100 random starting points per maturity, only a single start converged in each case, and in every instance the estimated score coefficients $A_{ij}$ collapsed to exactly zero, reducing the model to the constant-probability specification. The remaining 99 starts with non-zero initial $A_{ij}$ values uniformly failed to converge, suggesting that the GAS likelihood surface is essentially flat or ill-conditioned in the score-coefficient dimensions for these yield curve series. These symptoms are consistent with the identifiability issue diagnosed in the Monte Carlo study (Section~\ref{sec:sim_results}), where a ridge in the joint $(\sigma^2, A)$ likelihood drove $\hat A$ to zero regardless of the true parameter.

Two features of the present application plausibly make the collapse so systematic. First, the ridge is tied to the chosen scaling of the score: with $S_{ij,t}$ equal to the square root of the inverse Fisher information, the magnitude of the scaled score depends on the regime variances, so an increase in $A_{ij}$ can be offset by small adjustments of $\sigma_i^2$ (and vice versa) while leaving the filtered probabilities and the likelihood nearly unchanged. On real data, where no specification is exactly true, this flat direction gives the optimizer no gradient to follow away from $A=0$. Second, the fit comparison below shows that the regime information in these series is carried by the slowly moving yield \emph{level}, whereas the score is a function of the one-step-ahead prediction error of yield \emph{changes}. A plausible reading is that, conditional on the regime intercepts, this error behaves largely like noise, leaving the score-driven update little systematic signal to exploit in the first place. We therefore exclude the GAS specification from the results tables below and focus our comparison on the Constant, TVP, and Exogenous models. Possible remedies for the identifiability problem, such as alternative scaling choices or penalized estimation, are discussed in Section~\ref{sec:conlusions}.

Table~\ref{tab:results_summary} reports the model fit statistics for each maturity and model specification. Table~\ref{tab:param_estimates} presents the estimated regime-specific parameters.

\begin{table}[!htbp]
\centering
\caption{Model fit comparison across maturities and model specifications. $K=3$ regimes with regime-specific means and variances. $\ell$ denotes the maximized log-likelihood, $p$ the number of parameters.}
\label{tab:results_summary}
\begin{tabular}{ll rrr r}
\toprule
Maturity & Model & $\ell$ & AIC & BIC & $p$ \\
\midrule
\multirow{3}{*}{1\,m}
  & Constant  & $-8.98$   & $41.97$  & $95.91$  & 12 \\
  & TVP       & $-7.43$   & $50.86$  & $131.77$ & 18 \\
  & Exogenous & $12.42$   & $11.16$  & $92.07$  & 18 \\
\midrule
\multirow{3}{*}{12\,m}
  & Constant  & $-111.35$ & $246.71$ & $300.65$ & 12 \\
  & TVP       & $-108.89$ & $253.78$ & $334.69$ & 18 \\
  & Exogenous & $-86.77$  & $209.55$ & $290.46$ & 18 \\
\midrule
\multirow{3}{*}{36\,m}
  & Constant  & $-191.75$ & $407.50$ & $461.45$ & 12 \\
  & TVP       & $-191.07$ & $418.13$ & $499.05$ & 18 \\
  & Exogenous & $-180.00$ & $396.00$ & $476.91$ & 18 \\
\midrule
\multirow{3}{*}{72\,m}
  & Constant  & $-165.14$ & $354.28$ & $408.22$ & 12 \\
  & TVP       & $-164.77$ & $365.55$ & $446.46$ & 18 \\
  & Exogenous & $-155.03$ & $346.05$ & $426.97$ & 18 \\
\bottomrule
\end{tabular}
\end{table}

\begin{table}[!htbp]
\centering
\caption{Estimated regime-specific parameters. $\hat{\mu}_i$ and $\hat{\sigma}_i^2$ denote the estimated mean and variance for regime $i$, respectively. Regimes are ordered by increasing variance.}
\label{tab:param_estimates}
\begin{tabular}{ll rrr rrr}
\toprule
 & & \multicolumn{3}{c}{Mean ($\hat{\mu}_i$)} & \multicolumn{3}{c}{Variance ($\hat{\sigma}_i^2$)} \\
\cmidrule(lr){3-5} \cmidrule(lr){6-8}
Maturity & Model & Regime 1 & Regime 2 & Regime 3 & Regime 1 & Regime 2 & Regime 3 \\
\midrule
\multirow{3}{*}{1\,m}
  & Constant  & $-0.0006$ & $0.0283$  & $-0.1140$ & $0.0007$ & $0.0565$ & $1.1602$ \\
  & TVP       & $-0.0005$ & $0.0290$  & $-0.1180$ & $0.0007$ & $0.0569$ & $1.1700$ \\
  & Exogenous & $0.0081$  & $0.0281$  & $-0.1782$ & $0.0028$ & $0.0833$ & $1.6044$ \\
\midrule
\multirow{3}{*}{12\,m}
  & Constant  & $0.0116$  & $-0.0139$ & $-0.0176$ & $0.0109$ & $0.1580$ & $1.8673$ \\
  & TVP       & $0.0089$  & $-0.0093$ & $-0.0438$ & $0.0108$ & $0.1556$ & $1.8427$ \\
  & Exogenous & $0.0077$  & $0.0047$  & $-0.1142$ & $0.0031$ & $0.1065$ & $1.5596$ \\
\midrule
\multirow{3}{*}{36\,m}
  & Constant  & $0.0287$  & $-0.0238$ & $0.0302$  & $0.0240$ & $0.1407$ & $1.0358$ \\
  & TVP       & $0.0279$  & $-0.0236$ & $0.0302$  & $0.0243$ & $0.1411$ & $1.0487$ \\
  & Exogenous & $0.0245$  & $-0.0143$ & $-0.0346$ & $0.0208$ & $0.1338$ & $0.8245$ \\
\midrule
\multirow{3}{*}{72\,m}
  & Constant  & $0.0009$  & $-0.0189$ & $0.0696$  & $0.0479$ & $0.1415$ & $0.6037$ \\
  & TVP       & $-0.0010$ & $-0.0157$ & $0.0620$  & $0.0473$ & $0.1413$ & $0.6003$ \\
  & Exogenous & $0.0260$  & $-0.0151$ & $-0.0228$ & $0.0295$ & $0.1052$ & $0.3791$ \\
\bottomrule
\end{tabular}
\end{table}

\begin{figure}[!htbp]
\centering
\begin{subfigure}[b]{0.48\textwidth}
  \includegraphics[width=\textwidth]{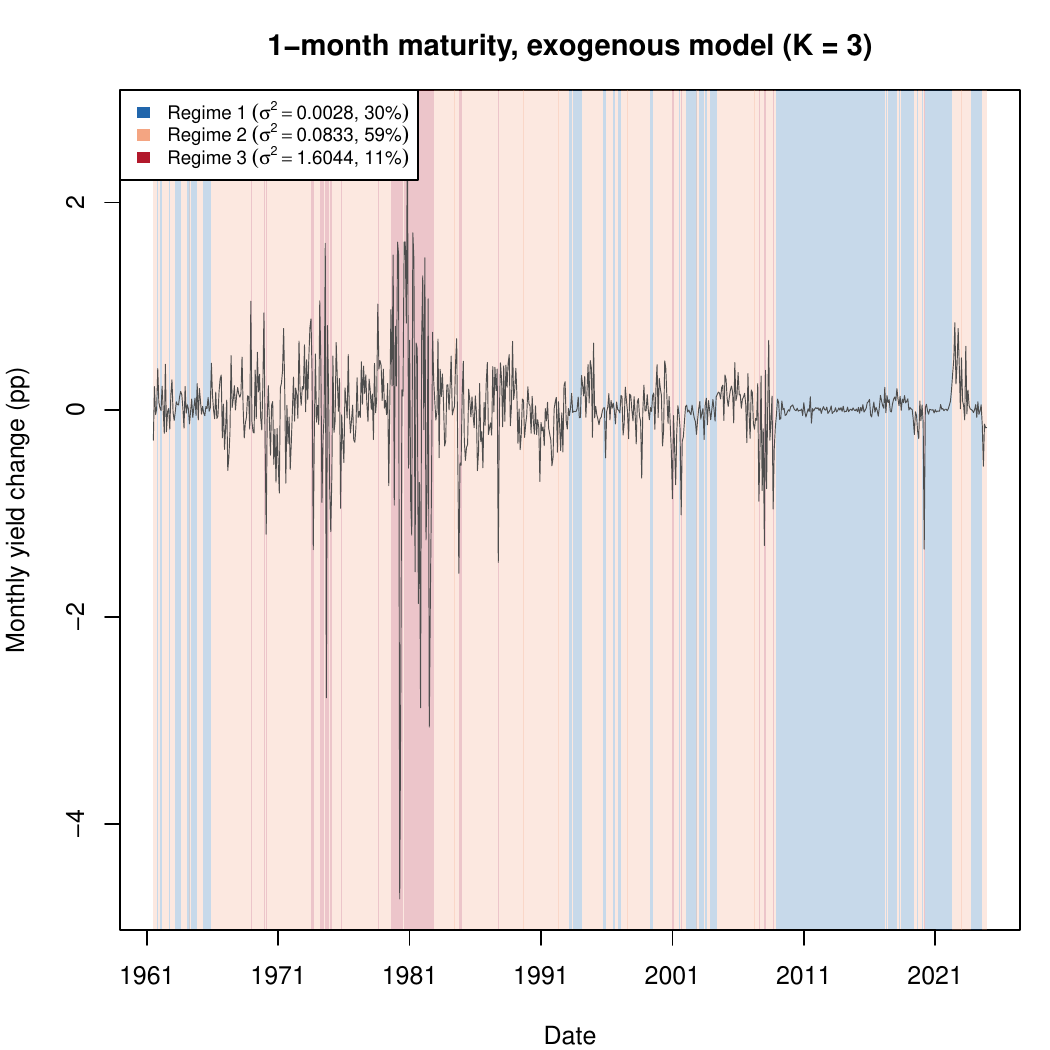}
  \caption{1-month maturity}
\end{subfigure}
\hfill
\begin{subfigure}[b]{0.48\textwidth}
  \includegraphics[width=\textwidth]{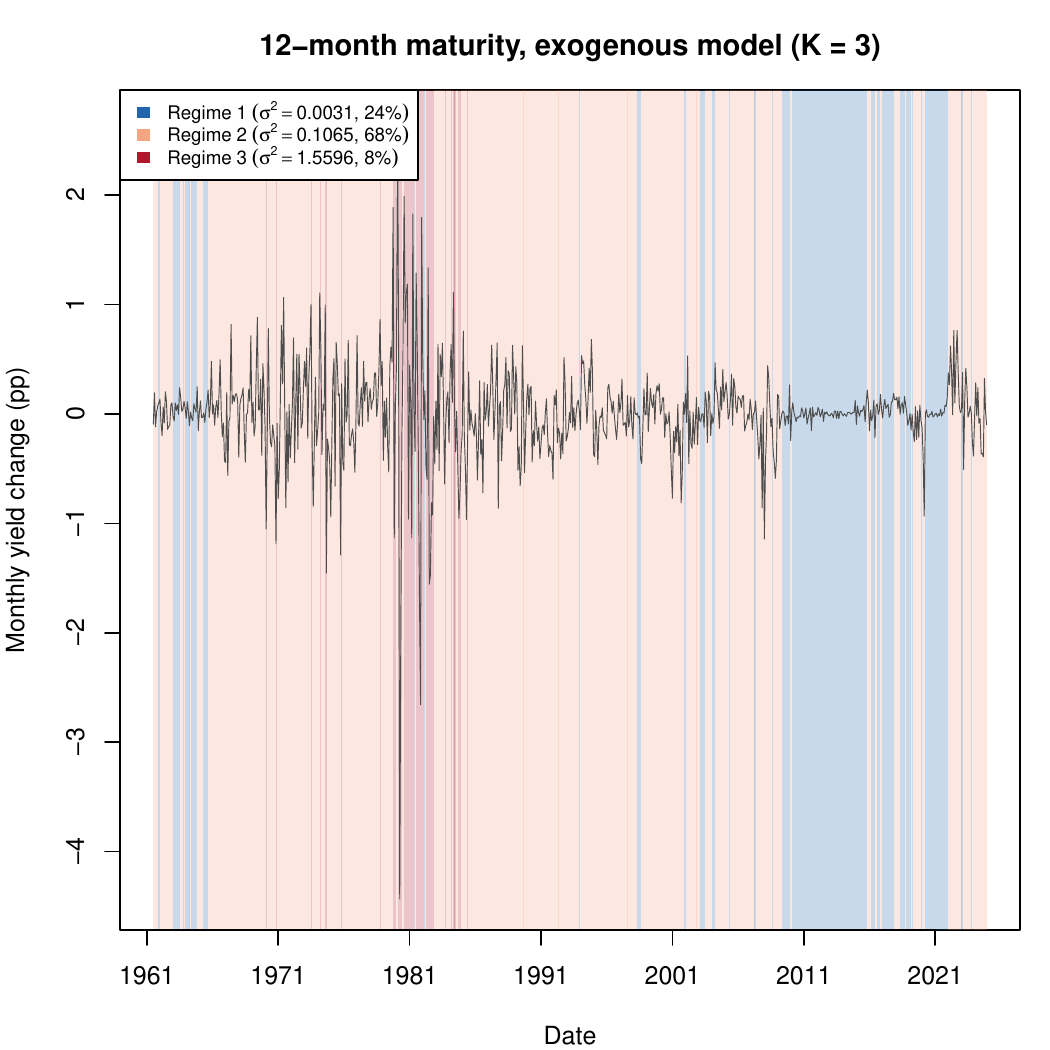}
  \caption{12-month maturity}
\end{subfigure}

\vspace{0.5em}

\begin{subfigure}[b]{0.48\textwidth}
  \includegraphics[width=\textwidth]{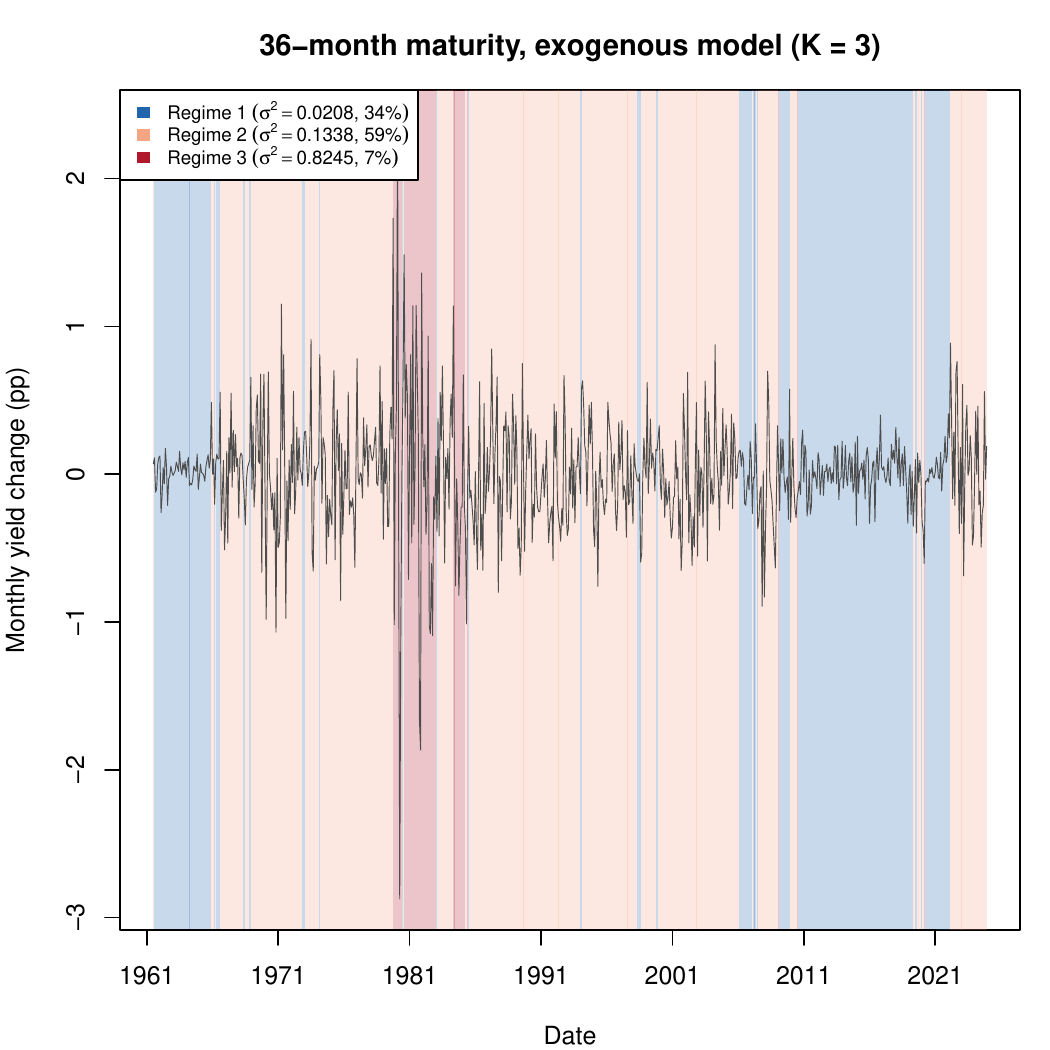}
  \caption{36-month maturity}
\end{subfigure}
\hfill
\begin{subfigure}[b]{0.48\textwidth}
  \includegraphics[width=\textwidth]{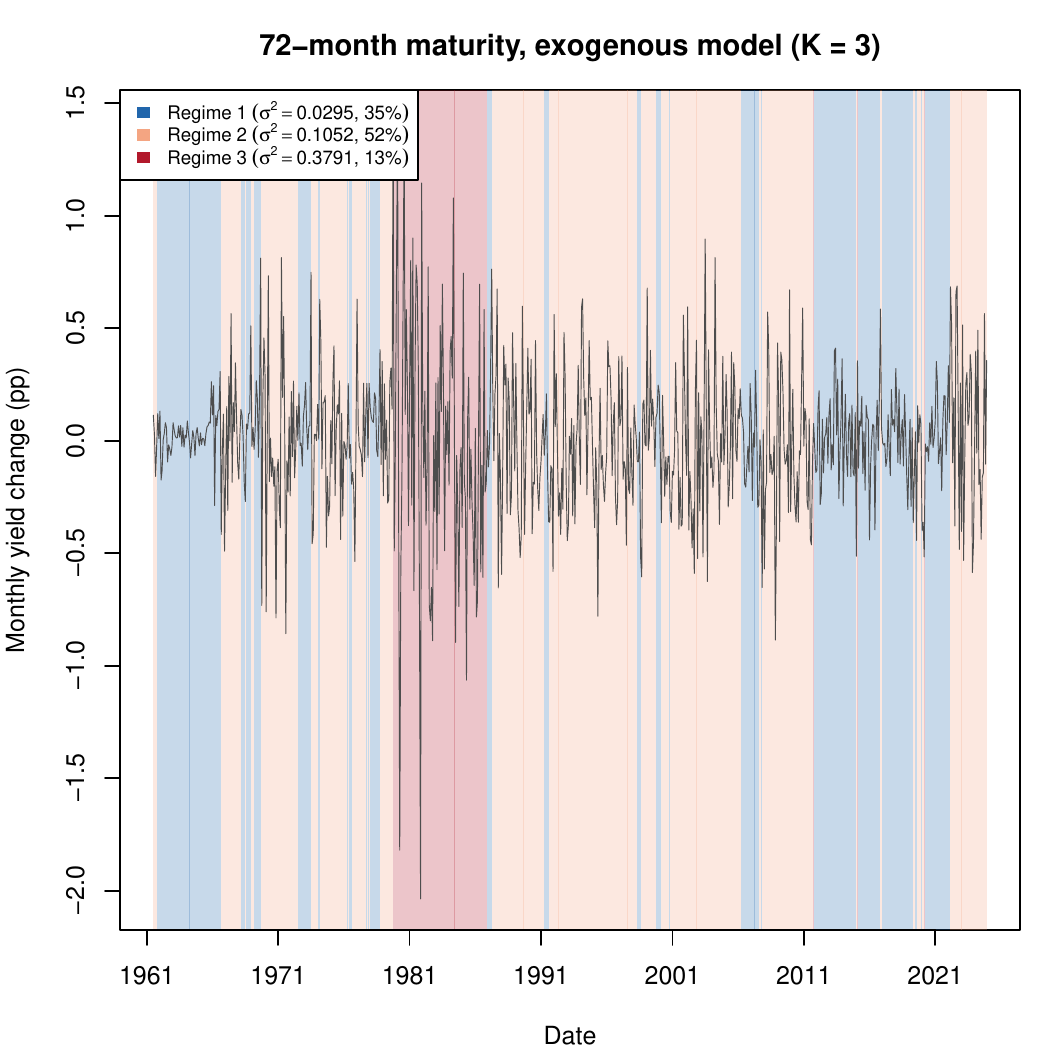}
  \caption{72-month maturity}
\end{subfigure}
\caption{Filtered regime classifications from the Exogenous model ($K=3$) for four maturities. Background shading indicates the most probable regime at each time point, ordered by variance: blue = low volatility, salmon = moderate volatility, red = high volatility.}
\label{fig:regime_plots}
\end{figure}

Figure~\ref{fig:regime_plots} displays the filtered regime classifications from the Exogenous model across all four maturities. At each point in time, the background shading indicates the most probable regime, with regimes ordered by variance: blue corresponds to the low-volatility regime, salmon to moderate volatility, and red to the high-volatility regime.

\subsection{Economic interpretation}\label{subsec:econ-interp}

The filtered regime classifications broadly track major macroeconomic episodes, though regime shifts are not synchronous across maturities---since each maturity is filtered independently, shifts are staggered in timing and sharpness. 
The low-volatility regime dominates the early sample through the mid-1960s, consistent with the stable monetary environment under Bretton Woods. Volatility rises to the moderate regime over the mid-to-late 1960s, coinciding with the onset of the Great Inflation \citep{meltzer2005origins}---Vietnam-era fiscal pressures, the 1971 breakdown of Bretton Woods, and the 1973 oil shock. The 1973 shock imprints mainly on the 1-month maturity; we read it as evidence that short and long ends load on different risk factors under acute supply shocks: the short end tracks policy and near-term inflation risk, the long end is anchored by expectations and term premia. This staggering is quantifiable: the short end (1m, 12m) is already switching noisily between the low- and moderate-volatility regimes from the very start of the sample and settles into the moderate regime by late 1965, whereas the long end (36m, 72m) remains cleanly low-volatility until a sharp, sustained shift around August--September 1966, roughly a year later. A similar lag appears later in the sample: after the 2008 crisis, the short end re-enters the low-volatility regime within months (1m in December 2008; 12m in May 2009). The 72-month series, by contrast, does not settle back into it until October 2011.

The high-volatility regime is most prominently active in the early 1980s, coinciding with the Volcker disinflation \citep{goodfriend2005incredible} and the 1979 oil shock, when the shift to reserve targeting drove short rates above $20\%$. Its magnitude falls sharply with maturity---the high-volatility variance drops from $\sigma^2\approx 1.60$ (1m) to $\sigma^2\approx 0.38$ (72m)---so every maturity enters the turbulent regime but the long end does so far more mildly, again consistent with term-premium smoothing. 

The mid-1980s through 2008 are classified predominantly as moderate volatility, consistent with the Great Moderation \citep{stock2003has}. The 2004--06 Greenspan conundrum \citep{backus2007cracking} is visible in this window, though not as the simple ``short turns turbulent, long stays calm'' split one might expect: no maturity ever enters the high-volatility regime during this episode. Instead, the short end (1m, 12m)---which had dipped back into the low-volatility regime during 2002 through early 2004---steps up to the moderate regime almost exactly when the Fed began tightening in mid-2004 (1m in June 2004; 12m in April 2004) and remains there. The long end (36m, 72m), continuously in the moderate regime since 2000, instead drops into the low-volatility regime precisely during the heart of the conundrum (36m in February 2006--January 2007; 72m in April 2006--July 2007), even as short rates kept rising. The model thus registers the conundrum's decoupling of short and long yields, but as a divergence between the moderate and calm regimes rather than between turbulent and calm.

After 2008 the low-volatility regime re-emerges and persists into the early 2020s, reflecting the zero-lower-bound era and the Federal Reserve's large-scale asset purchases (quantitative easing, QE), though not uniformly. Brief, single-month excursions into the moderate or high-volatility regime punctuate the period at every maturity, scattered between 2008 and 2015. The 2015--18 hiking cycle leaves a longer, maturity-dependent mark, most visible at the 12-month maturity, which spends roughly a quarter of that window in the moderate regime, while the 36-month maturity is untouched throughout. The model measures the absolute variance of monthly yield changes (percentage points), so with the policy rate pinned near zero, absolute movements were compressed even when relative uncertainty was extreme. March 2020 is the sharpest test of this low-volatility backdrop, and it is not absorbed into it: the model classifies March 2020 itself as high-volatility at 1m, 36m, and 72m (filtered probability $>99\%$) and moderate-volatility at 12m ($\approx99\%$). The spike is brief, preceded by a February ramp into the moderate regime and followed by an immediate return to low volatility, with the filter back in the low-volatility regime by April 2020. In the underlying data, the pandemic shock produces a large negative yield-change spike, largest at the short end and weaker at 36/72m: $-1.34$, $-0.94$, $-0.60$, and $-0.52$ percentage points at 1m, 12m, 36m, and 72m, respectively. The regime classification does not simply mirror this magnitude ordering: the 36m and 72m series are tagged high-volatility despite smaller raw moves than 12m, because each maturity's regimes are calibrated to its own historical volatility scale.

On fit, the exogenous (lagged-level) specification attains the lowest AIC at all four maturities. On BIC it wins at the short end (1m, 12m) but not at 36 and 72 months, where the constant model's lower parameter count prevails. This gradient is itself interpretable: the value of level-driven transition dynamics concentrates where the short-rate level most strongly governs volatility. Notably, the lagged-change (TVP) model never improves on the constant model, whereas the lagged-level model does at the short end---it is specifically the level, not recent dynamics, that carries regime information. This is the discrete-regime analogue of level-dependent short-rate volatility (the CKLS class, \citealp{ckls1992}, where conditional volatility scales with the rate level): high levels raise the probability of the high-variance regime, information neither the constant nor the TVP model exploits. The closest antecedent is \citet{gray1996modeling}, whose generalized regime-switching model of the short rate already lets the switching probabilities depend on the lagged rate level through a probit link. Our results extend that mechanism from two regimes to three with regime-specific variances, and from the short rate to the maturity spectrum, where the maturity gradient in the AIC/BIC ranking shows the value of level-driven transition dynamics concentrating at the short end. A complementary finding is reported by \citet{vanos2024accelerating}, who allow the transition probabilities of a dynamic factor Markov-switching model of U.S.\ business cycle turning points to vary with the term spread and find that this yield-curve information materially accelerates real-time peak detection. That two quite different regime-switching frameworks---a business-cycle factor model driven by the yield spread and our univariate volatility-regime model driven by the yield level---both identify yield-based variables as the operative TVTP driver suggests that the information content of interest-rate levels and spreads for regime dynamics is a robust feature of the data rather than an artefact of either specification.
The recovery of these documented episodes indicates the regimes are economically sensible. We stress, however, that episode-matching would be reproduced by any volatility-regime model, so it corroborates the plausibility of the regimes, not the value of the TVTP structure---the latter rests on the fit comparison above.

\section{Conclusions and future work}\label{sec:conlusions}

\noindent The paper studies time-varying transition probability (TVTP) Markov-switching (MS) models with various dynamics in the transition probability matrix. The regime switching models are univariate Markov-switching models with $K$ regimes, each with its own mean and variance and regime changes follow a first‑order Markov chain. A unified R package, \texttt{multiregimeTVTP}, is developed to simulate data, and estimate the model parameters for arbitrary $K\geq 2$.  Monte Carlo simulation shows that regime means/variances and average transition probabilities are well estimated, but TVTP coefficients---especially in GAS models and three‑regime settings---are hard to identify. The misspecification analysis shows that the one-step-ahead conditional mean is insensitive to the choice of TVTP specification, whereas the filtered regime probabilities are accurately recovered only under correct specification. The practical value of the TVTP structure therefore lies in characterizing the regime dynamics: the latent-state probabilities and the timing of regime changes respond to the transition specification, while the conditional mean does not.

In the yield curve application, the level‑driven specification attains the lowest AIC at all four maturities and the lowest BIC at the short end of the curve. This pattern suggests that, for these data, regime variation is closely linked to the level of interest rates, which is relevant for model specification and selection in yield-curve applications. Apart from more empirical applications of the TVTP-MS model with more than two regimes, several extensions merit further work. The non-identifiability of the GAS score coefficient suggests concrete remedies to investigate. Since the $(\sigma^2, A)$ ridge arises from the inverse-Fisher scaling of the score, which makes the magnitude of the scaled score depend on the regime variances, a simpler scaling choice (e.g.\ the identity matrix) could decouple the score coefficient from the variances and restore identifiability, at the cost of a score whose variance is no longer standardized. Alternatively, penalized maximum likelihood could regularize $A$ along the flat direction of the likelihood, and reparameterizations that fix $\sigma^2$ or constrain the product of $A$ and the score scale would break the ridge directly. Further, evaluating the score-driven specification by its probability-tracking performance, as in \citet{bazzi2017time}, rather than by coefficient recovery, is a complementary route. Beyond the GAS specification, transition probabilities could be cyclical \citep{bazzi2017time}, and the impact of the specific functional form mapping $f_{ij,t}$ to the transition probabilities (e.g.\ the element-wise logistic link used here versus a multinomial softmax link) remains an interesting question for future work.
\section*{Data availability statement}
\noindent The U.S.\ Treasury zero-coupon yield data of \citet{liu2021reconstructing} are publicly available from the authors' website (\url{https://sites.google.com/view/jingcynthiawu/yield-data}); the monthly series used in this paper are supplied as Supporting Information. The R package \texttt{multiregimeTVTP}, together with the scripts reproducing all simulation and empirical results, is available at \url{https://github.com/smodee/multiregime-TVTP}.

\section*{Conflict of interest statement}
\noindent The authors declare no conflict of interest.

\section*{Funding statement}
\noindent This research received no specific grant from any funding agency in the public, commercial, or not-for-profit sectors.

\bibliography{source}
\end{document}